\documentclass[twocolumn]{aastex631}

\pdfoutput=1 %for arXiv submission
\usepackage{amsmath,amstext}
\usepackage[T1]{fontenc}
\usepackage{apjfonts} 
\usepackage[figure,figure*]{hypcap}
\usepackage{fixltx2e}

\def\deg{^\circ} 
\newcommand{\x}{\sout}

\newcommand{\HI}{H\,{\sc i}}
\newcommand{\DM}{\mathrm{DM}}
\newcommand{\Hp}{\mathrm{H}^{+}}

\newcommand{\Nwnm}{N_{{H},i}^{\text{WNM}_\text{\HI}}}
\newcommand{\Nwim}{N_{{H},i}^{\text{WIM}}}
\newcommand{\XHwim}{\chi_{H}^{\text{WIM}}}
\newcommand{\XHwnm}{\chi_{H}^{\text{WNM}_\text{\HI}}}
\newcommand{\XHewim}{\chi_{He}^{\text{WIM}}}

\renewcommand{\vec}[1]{\mathbf{#1}} 

\begin{document}

\title{All We Are Is Dust In The WIM: Constraints on Dust Properties in the Milky Way's Warm Ionized Medium}

\correspondingauthor{J. L. West et al.}
\email{jennifer.west@nrc-cnrc.gc.ca}

\author[0000-0001-7722-8458]{J. L. West}
\affil{National Research Council Canada, Herzberg Research Centre for Astronomy and Astrophysics, Dominion Radio Astrophysical Observatory, PO Box 248, Penticton, BC V2A 6J9, Canada}
\affil{Dunlap Institute for Astronomy and Astrophysics, University of Toronto, 50 St. George Street, Toronto, ON M5S 3H4, Canada}

\author[0000-0002-3382-9558]{B. M. Gaensler}
\affil{Dunlap Institute for Astronomy and Astrophysics, University of Toronto, 50 St. George Street, Toronto, ON M5S 3H4, Canada}
\affil{Department of Astronomy and Astrophysics, University of Toronto, 50 St. George Street, Toronto, ON M5S 3H4, Canada}

\affil{Division of Physical and Biological Sciences, University of California Santa Cruz, Santa Cruz, CA 95064, USA}

\author{M.-A. Miville-Desch\^enes}
\affil{Institut d'Astrophysique Spatiale, CNRS, Univ. Paris-Sud, Universit\'e Paris-Saclay, B\^at. 121, 91405 Orsay Cedex, France}

\author[0000-0002-6317-3190]{N. Mahajan}
\affil{Department of Astronomy and Astrophysics, University of Toronto, 50 St. George Street, Toronto, ON M5S 3H4, Canada}
\affil{Dunlap Institute for Astronomy and Astrophysics, University of Toronto, 50 St. George Street, Toronto, ON M5S 3H4, Canada}

\author{J. Dechant}
\affil{Dunlap Institute for Astronomy and Astrophysics, University of Toronto, 50 St. George Street, Toronto, ON M5S 3H4, Canada}
\affil{Department of Physics and Astronomy, University of Calgary, 2500 University Drive NW
Calgary, AB T2N 1N4, Canada}
\affil{Athabasca University, 1 University Dr, Athabasca, AB T9S 3A3}

\author{F. Boulanger}
\affil{Institut d'Astrophysique Spatiale, CNRS, Univ. Paris-Sud, Universit\'e Paris-Saclay, B\^at. 121, 91405 Orsay Cedex, France}

\author{P. G. Martin}
\affil{Canadian Institute for Theoretical Astrophysics, University of Toronto, 60 St. George Street, Toronto, ON M5S 3H8, Canada}

\author[0000-0002-7588-976X]{I. A. Zelko}
\affil{Canadian Institute for Theoretical Astrophysics, University of Toronto, 60 St. George Street, Toronto, ON M5S 3H8, Canada}

\begin{abstract}
We present a comparison of the presence and properties of dust in two distinct phases of the Milky Way's interstellar medium: the warm neutral medium (WNM) and the warm ionized medium (WIM). Using distant pulsars at high Galactic latitudes and vertical distance ($|b| > 40\deg$, $D \sin|b| > 2 \mathrm{\,\, kpc}$) as probes, we measure their dispersion measures and the neutral hydrogen component of the warm neutral medium ($\text{WNM}_\text{\HI}$) using \HI\ column density. Together with dust intensity along these same sightlines, we separate the respective dust contributions of each ISM phase in order to determine whether the ionized component contributes to the dust signal. We measure the temperature ($T$), spectral index ($\beta$), and dust opacity ($\tau/N_{H}$) in both phases. We find $T~{\text{(WNM}_\text{\HI})}=20^{+3}_{-2}$~K, $\beta~{\text{(WNM}_\text{\HI})} = 1.5\pm{0.4}$, and $\tau_{\text{353}}/N_{H}~{\text{(WNM}_\text{\HI})}=(1.0\pm0.1)\times 10^{-26}$~cm$^2$. Assuming that the temperature and spectral index are the same in both the WNM$_\text{\HI}$ and WIM, and given our simple model that widely separated lines-of-sight can be fit together, we find evidence that there is a dust signal associated with the ionized gas and $\tau_{\text{353}}/N_{H}~\text{(WIM)}=(0.3\pm0.3)\times 10^{-26}$, which is about three times smaller than $\tau_{\text{353}}/N_{H}~{\text{(WNM}_\text{\HI})}$. We are 80\% confident that $\tau_{\text{353}}/N_{H}~\text{(WIM)}$ is at least two times smaller than $\tau_{\text{353}}/N_{H}~{\text{(WNM}_\text{\HI})}$. 
\end{abstract}


\section{Introduction} 
\label{sec:intro}

Dust plays an important role in the interstellar medium (ISM) : it is a powerful catalyst for astrochemistry, is the dominant source of opacity at optical wavelengths, and couples neutral gas to magnetic fields \citep{draine2011c, saintonge2022, mcclure-griffiths2023}. However, the detailed life-cycle of dust grains in the ISM is not yet understood, nor do we fully know how the properties of dust vary across different phases of the ISM. In this paper, we consider the properties of dust in two pervasive phases of the Galactic ISM: the warm ionized medium (WIM) and the neutral hydrogen component of the warm neutral medium ($\text{WNM}_\text{\HI}$\footnote{We add the subscript \text{\HI} here to remind the reader where we are specifically referring to the neutral atomic hydrogen component of the WNM, as opposed to the total WNM that may also include molecular hydrogen. Elsewhere in the text we do not include the subscript when discussing the generalized WNM phase.}). 

This warm ionized medium (WIM) consists of warm ($6000 - 10000 \,\, \mathrm{K}$), low density ($n_e = 0.03 - 0.08 \,\, \mathrm{cm^{-3}}$), highly ionized \citep[with fractional ionization of hydrogen, $\chi_\text{H} > 0.7$, ][]{Rey98} gas with an average volume filling fraction of $f \approx 0.2-0.4$ within a 2-3 kpc thick layer around the Galactic midplane \citep{H09}. At the midplane, however, the filling fraction is $f_0 < 0.1$ implying that the WIM is a much less significant part of the ISM at the midplane, but at intermediate vertical heights, it becomes a dominant phase. The WIM has a vertical exponential scale height of 1.5--1.8~kpc \citep{G08, 2020ApJ...897..124O}. The other ISM phase that makes a significant contribution to most sightlines is the WNM, a neutral analogue ($\chi_\text{H} \approx 0.1$) of the WIM, with both phases having similar temperatures and densities. \cite{Howk03} show that the WIM and the WNM are spatially and kinematically correlated. There are very few regions in the sky where one might find only the WNM or only the WIM. However, the FWHM thickness of the neutral \HI\ layer is only $\sim230$~pc \citep{Ferr}, much less than the vertical scale height of the WIM.

The presence of dust throughout the WNM has been long established \citep[e.g.,][]{Bou96}. The dust opacity relates the total gas column density, $N_{H}$, to the total dust optical depth, $\tau$, along any sight-line. \cite{planck2013-p06b} compared thermal dust and \HI\ emission over the entire sky and found an opacity $\tau_{353}/N_{H} \approx 6.3 \times10^{-27}$~cm$^2$ for the range $1.2 \times 10^{20} < N_{\rm H\,I} < 2.5 \times10^{20}$~cm$^{-2}$, where $\tau_{353}$ is the optical depth of dust at 353~GHz and $N_{\rm H\,I}$ is the column density of \HI. \cite{planck2013-XVII} used the dust-\HI\ correlation to measure the properties of dust in the diffuse interstellar medium. They find $\tau_{353} / N_{H} = (7.1 \pm 0.6) \times 10^{-27}$~cm$^2$, the spectral index, $\beta=1.53\pm0.03$, and temperature, $T=19.9\pm1.0$~K.

Considerably less is known about dust in the WIM.  Initial studies only obtained upper limits on the dust contribution of the WIM \citep{1988ApJ...330..964B, Bou96,1998ApJ...508...74A}. The first clear detection of dust in the WIM was by \cite{L99,L00}. \cite{L99} identified excess far-infrared dust emission seen by {\it COBE} that was uncorrelated with \HI, which they associated with the WIM. \cite{L00} jointly correlated the spatial distribution of dust with that of \HI\ (which they associated with the WNM, and we designate as WNM$_\text{\HI}$) and H$\alpha$ (associated with the WIM) at high Galactic latitudes, and found that both the WNM$_\text{\HI}$ and WIM contribute to the {\it COBE}\ dust signal. \cite{L00} used a $\nu^2$ modified {\it Planck} curve to model dust emission associated with both \HI\ and $\Hp$ gas. For \HI\ gas, they found a dust temperature of $17.2  \,\, \mathrm{K}$ and a WNM$_\text{\HI}$ opacity $\tau / N_{\rm H\,I} = 8.3 \times 10^{-26} (\lambda / 250 \mathrm{\mu m})^{-2} \mathrm{cm}^{2}$.~\footnote{Here we include the subscript \text{\HI} in $\tau / N_{\rm H\,I}$ since that is the notation used by \cite{L00}. They use this notation to distinguish the opacity for neutral and $\Hp$ components.} For $\Hp$ gas, \cite{L00} inferred a dust temperature in the range 16--18~K; assuming that \HI\ gas has the same dust temperature as $\Hp$ gas, they derive a WIM opacity of $\tau / N_{\Hp} = 1.1 \times 10^{-25} (\lambda / 250 \mathrm{\mu m})^{-2} \mathrm{cm}^{2}$, implying that the dust abundance in the WIM is comparable to that in the WNM. They assumed that both dust components contain the same type of dust, that is their spectral energy distributions (SED) have the same shapes.

\citet{Howk03, 2006ApJ...637..333H} and \citet{howk2012} analyzed UV spectra of bright stars in the globular clusters M3 and M5, to carefully decompose the gas phase contributions of the WNM, WIM, and also the hot ionized medium (HIM) towards these sight-lines. \citet{Howk03} found evidence for the presence of dust in both the WIM and WNM phases of the ISM suggesting ``similar dust destruction and formation histories''. \citet{2006ApJ...637..333H} and \citet{howk2012} focused on the difference between warm and hot phases. In these analyses they find that the warm phase contains $\sim80\%$ of the total ionized gas column, and the hot phase $<20\%$

There is also qualitative evidence for the presence of dust in the WIM. In a multi-wavelength study of the Eridanus Superbubble, \citet{1999ASPC..168..211H} found evidence that all the phases of the interstellar medium, including the WIM and WNM, co-exist in close proximity. \citet{2018A&A...615L...3J} found an alignment between several different tracers of the interstellar medium including neutral hydrogen, synchrotron emission, and emission from polarized dust, which is also evidence of the co-existence of ionized and neutral media.

In this paper, we aim to use the latest \HI\ surveys and accurate electron column densities from pulsar dispersion measures (DMs) to decompose sightlines into the $\text{WNM}_\text{\HI}$ and the WIM. Then, using the intensities from thermal dust emission to those same sightlines from {\it Planck} at 353, 545, and 857 GHz, as well as 3000 GHz (100 $\mu$m) data from Improved Reprocessing of the {\it Infrared Astronomical Satellite} (IRIS), we determine dust parameters associated with the two ISM phases. The goal is to determine whether dust is pervasive in the WIM, and then to investigate whether and in what ways the WIM-correlated dust parameters differ from the $\text{WNM}_\text{\HI}$-correlated dust. 

In order to make this work possible, we must assume that the dust emission per H atom are the same across the sky at each frequency. This simplification is required to perform the analysis but it ignores observational 
evidence for variations of the dust emission per H atom over the sky \citep[e.g.,][]{planck2013-XVII}. 
Variations in dust temperature and dust-to-gas ratio can both contribute, as well as variations in the proportion of cold neutral and warm neutral gas 
\citep{2019ApJ...874..171C}. The analysis does not require that the total amount of dust along each line-of-sight be consistent (i.e., the emission need not be proportional to distance).

We provide details of the data products in Sec.~\ref{sec:data}. Our analysis method is presented in Sec.~\ref{sec:analysis}, and discussion and conclusions are presented in Secs.~\ref{sec:discussion} and \ref{sec:conclusion}, resepectively.

\section{Data}
\label{sec:data}

\begin{deluxetable*}{lcccccccccccc}
\rotate
%\tabletypesize{\tiny}
\tablecaption {Gas and dust properties along selected sightlines. The top half of the table lists radio pulsars with parallaxes, while the bottom half lists globular clusters containing radio pulsars. In both cases, we require $|b| > 40^\circ$ and D $\sin (|b|) > 2$~kpc for a sightline to be included in this sample.\label{tc}}
\tablehead{\colhead{Object} &  \colhead{$\ell$ ($^\circ$)} & \colhead{$b$ ($^\circ$)} & \colhead{Distance} &  \colhead{D $\sin (|b|)$} &\colhead{$N_{\rm psr}$} & \colhead{$\DM$} & \colhead{$N_{\rm H\,I}$} & \colhead{$I_{353}$} & \colhead{$I_{545}$ } & \colhead{$I_{857}$ } &\colhead{$I_{3000}$} & Refs. 
\\
\colhead{} & \colhead{} & \colhead{} & \colhead{(kpc)} & \colhead{(kpc)} & \colhead{} & \colhead{(pc~cm$^{-3}$)} & \colhead{($10^{20} \,\, \mathrm{cm^{-2}}$)} & \colhead{(MJy/sr)} & \colhead{(MJy/sr)} & \colhead{(MJy/sr)} &\colhead{(MJy/sr)} & \colhead{}
}
\startdata
PSR B1541+09    & $17.81$   & $+45.78$  & $5.9^{+0.6}_{-0.5}$ & $4.2\pm0.4$ & 1 & $34.9758 \pm 0.0016$  & $2.99$  &		$	0.238	\pm	0.007	$ & $	0.719	\pm	0.010	$ & $	1.756	\pm	0.014	$ & $	2.060	\pm	0.012	$ & 1, 2 \\
PSR J2248--0101  & $69.26$   & $-50.62$  & $3.9^{+1.4}_{-0.6}$ & $3.0^{+1.1}_{-0.5}$  &  1 & $29.05 \pm 0.03$  & $5.69$  &		$	0.420	\pm	0.006	$ & $	1.352	\pm	0.010	$ & $	3.644	\pm	0.015	$ & $	4.204	\pm	0.016	$ & 3, 4 \\
PSR J2346--0609  & $83.80$   & $-64.01$  & $3.6^{+0.6}_{-0.3}$ & $3.3^{+0.5}_{-0.2}$ & 1 & $22.504 \pm 0.019$    & $2.60$   &		$	0.207	\pm	0.006	$ & $	0.626	\pm	0.009	$ & $	1.413	\pm	0.015	$ & $	1.365	\pm	0.016	$ & 3, 4 \\
PSR B0148--06    & $160.37$  & $-65.00$  & $4.6^{+2.5}_{-1.4}$ & $4.2^{+2.2}_{-1.3}$ &  1 & $25.66 \pm 0.03$  & $2.12$  & 		$	0.183	\pm	0.007	$ & $	0.530	\pm	0.011	$ & $	1.146	\pm	0.016	$ & $	1.281	\pm	0.019	$ & 3, 4 \\
PSR B0149--16    & $179.31$  & $-72.46$  & $2.3^{+1.6}_{-0.7}$ & $2.2^{+1.5}_{-0.7}$ & 1 & $11.92577 \pm 0.00004$& $1.22$  &		$	0.176	\pm	0.006	$ & $	0.498	\pm	0.010	$ & $	1.045	\pm	0.015	$ & $	0.876	\pm	0.017	$ & 3, 5 \\
PSR B1254--10    & $305.21$  & $+52.40$  & $7.1^{+13.2}_{-2.2}$ & $5.6^{+10.4}_{-0.8}$ & 1 & $29.634 \pm 0.009$    & $3.78$   &		$	0.296	\pm	0.007	$ & $	0.920	\pm	0.011	$ & $	2.380	\pm	0.020	$ & $	3.069	\pm	0.022	$ & 3, 4 \\
PSR J2129--5721  & $338.01$ & $-43.57$ & $3.6^{+5.0}_{-1.4}$ & $2.5^{+3.5}_{-1.0}$ & 1 & $31.8480 \pm 0.0004$ &  3.37 & 		$	0.246	\pm	0.005	$ & $	0.788	\pm	0.009	$ & $	1.948	\pm	0.018	$ & $	2.044	\pm	0.011	$ & 6, 7 \\ \hline
GC:M5           &  $3.86$    & $+46.80$  & $7.5\pm0.1$ & $5.45\pm0.04$ &  7 & $29.46 \pm 0.28$    & $3.12$ & 		$	0.245	\pm	0.007	$ & $	0.749	\pm	0.009	$ & $	1.826	\pm	0.012	$ & $	2.058	\pm	0.013	$ & 8, 9 \\
GC:M30          & $27.18$   & $-46.84$  & $8.5\pm0.1$ &  $6.17\pm0.07$ & 2 & $25.064 \pm 0.001$    & $3.35$ &		$	0.248	\pm	0.006	$ & $	0.779	\pm	0.009	$ & $	1.930	\pm	0.016	$ & $	2.347	\pm	0.016	$ & 8, 10 \\
GC:M3           & $42.22$   & $+78.71$  & $10.2\pm0.1$ & $9.98\pm0.08$ & 6 & $26.42 \pm 0.15$  & $1.04$ &		$	0.169	\pm	0.006	$ & $	0.474	\pm	0.009	$ & $	0.988	\pm	0.014	$ & $	0.644	\pm	0.016	$ & 8, 9 \\
GC:M13          & $59.01$   & $+40.91$  & $7.4\pm0.1$ & $4.86\pm0.05$ & 6 & $30.23 \pm 0.41$  & $1.31$    &		$	0.181	\pm	0.006	$ & $	0.530	\pm	0.010	$ & $	1.157	\pm	0.015	$ & $	0.840	\pm	0.012	$ & 8, 11 \\
GC:NGC~362 & $301.53$ & $-46.25$ & $8.8\pm0.1$ & $6.38\pm0.07$ & 6 &  $24.92 \pm 0.43$ & 2.42 &		$	0.222	\pm	0.003	$ & $	0.666	\pm	0.005	$ & $	1.523	\pm	0.007	$ & $	1.586	\pm	0.008	$ & 8, 12 \\
GC:47~Tuc      & $305.90$  & $-44.89$  & $4.52\pm0.03$ &  $3.19\pm0.02$ & 29    & $24.42 \pm 0.16$  & $2.80$  &		$	0.235	\pm	0.003	$ & $	0.700	\pm	0.006	$ & $	1.656	\pm	0.009	$ & $	1.483	\pm	0.008	$ & 8, 13 \\
GC:M53          & $332.96$  & $+79.76$  & $18.5\pm0.2$ & $18.2\pm0.2$ & 5 & $25.31 \pm 0.95$    & $1.51$  &		$	0.187	\pm	0.006	$ & $	0.533	\pm	0.009	$ & $	1.140	\pm	0.014	$ & $	0.987	\pm	0.017	$ & 8, 9 \\
\enddata
\tablecomments{This table lists the selected sightlines where there are high-latitude pulsars that sit outside most of the WIM. The ``GC:'' prefix indicates that the object is a globular cluster containing one or more pulsars. The fractional error of all \HI\ column densities is taken to be $3 \%$. The 9th through 12th columns list the intensity of dust emission at 353, 545, 857 and 3000~GHz, respectively.
Note that the position of PSR J2248--0101 falls in the 2\% of the sky not observed by IRAS, thus the 100 micron flux for this pulsar is from the Diffuse Infrared Background Experiment (DIRBE, 30$'$ resolution). References for distances and DMs:
(1)~\citet{2012ApJ...755...39V}
(2)~\citet{Bilous2016}
(3)~\citet{2019ApJ...875..100D}
(4)~\citet{2004MNRAS.353.1311H}
(5)~\citet{2015ApJ...808..156S}
(6)~\citet{2021MNRAS.507.2137R}
(7)~\citet{2022PASA...39...27S}
(8)~\citet{2021MNRAS.505.5957B}
(9)~\citet{2021ApJ...915L..28P} and references therein
(10)~\citet{2023ApJ...942L..35B} and references therein
(11)~\citet{2020ApJ...892...43W} and references therein
(12)~\citet{2016mks..confE...9S}; pulsar detections reported at \url{http://trapum.org/discoveries/}
(13) \citet{2021MNRAS.504.1407R} and references therein
}
\label{table:1}
\end{deluxetable*}

\subsection{Dispersion measures}
\label{ssec:dm}

For this analysis, we use radio pulsar $\DM$s to sample the entire electron column density of the WIM in the foreground to each pulsar. The $\DM$ of a pulsar is defined as,
\begin{equation}
\DM = \int_0^D n_e(s) \, ds
\end{equation}
where $D$ is the distance to the pulsar and $n_e(s)$ is the free electron density along the line element $ds$. The $\DM$ is thus simply a measure of the free electron column density along a sightline up to the pulsar. 

For the $\DM$s to correspond to the total free electron column density of the Milky Way along these sightlines, we must select pulsars that sit outside most of the WIM. Since the vertical scale height of the WIM is $\lesssim2$~kpc \citep{G08, 2020ApJ...897..124O}, high-latitude pulsars that lie above this scale height should satisfy this requirement. Therefore, we restrict our data to high-latitude sightlines ($|b| > 40\deg$) towards pulsars with vertical distances $D \sin|b| > 2 \,\, \mathrm{kpc}$. The cutoff of $|b| > 40\deg$ is selected to match that of \cite{G08}, and this value was chosen to avoid local features. We are thus required to look for pulsars with reliable distance estimates or at least reliable lower bounds for  distances.

\begin{figure*}[!ht]
\centering 
\begin{minipage}{8.5cm}
\includegraphics[width=8.4cm]{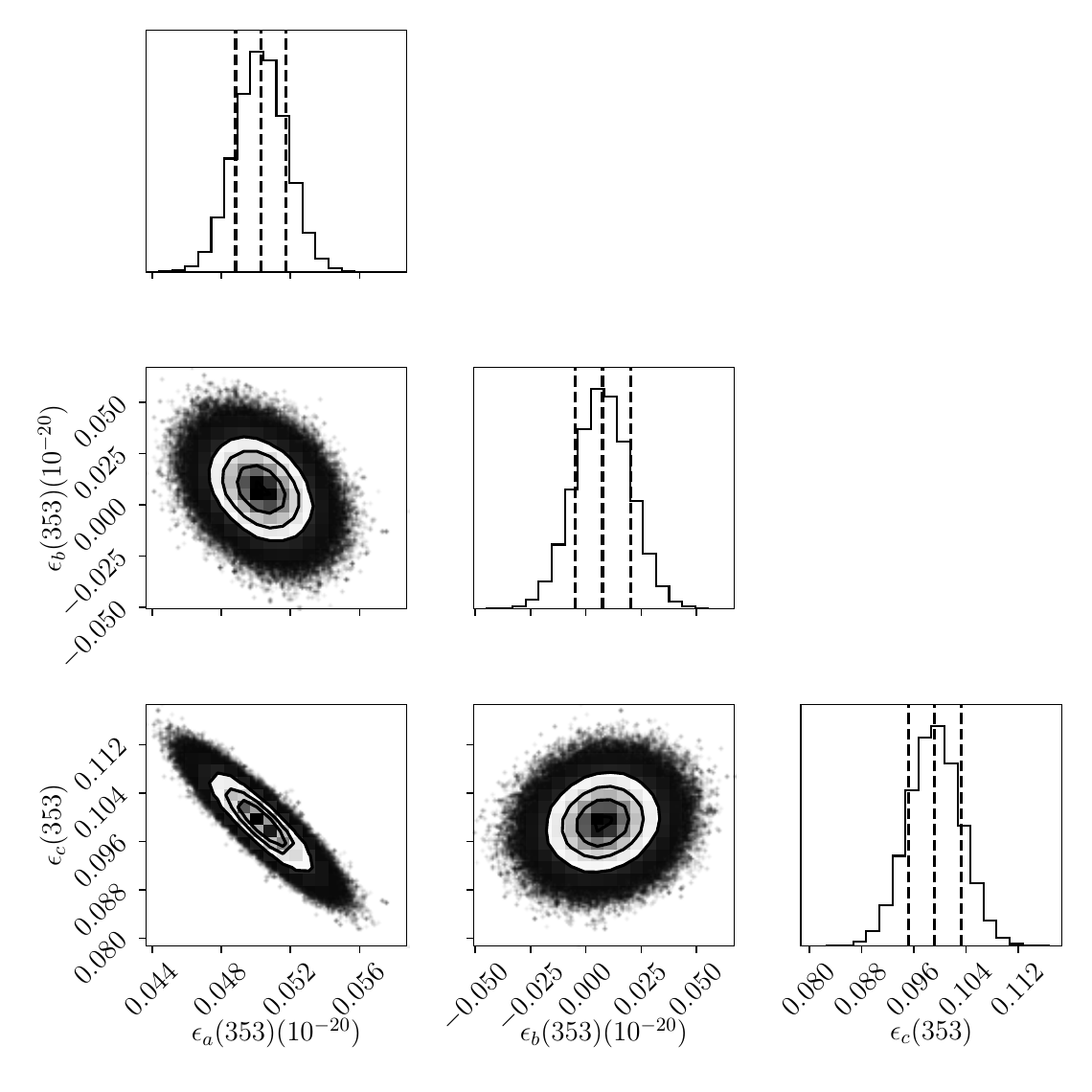}
\end{minipage}
\hfill
\begin{minipage}{8.5cm}
\includegraphics[width=8.4cm]{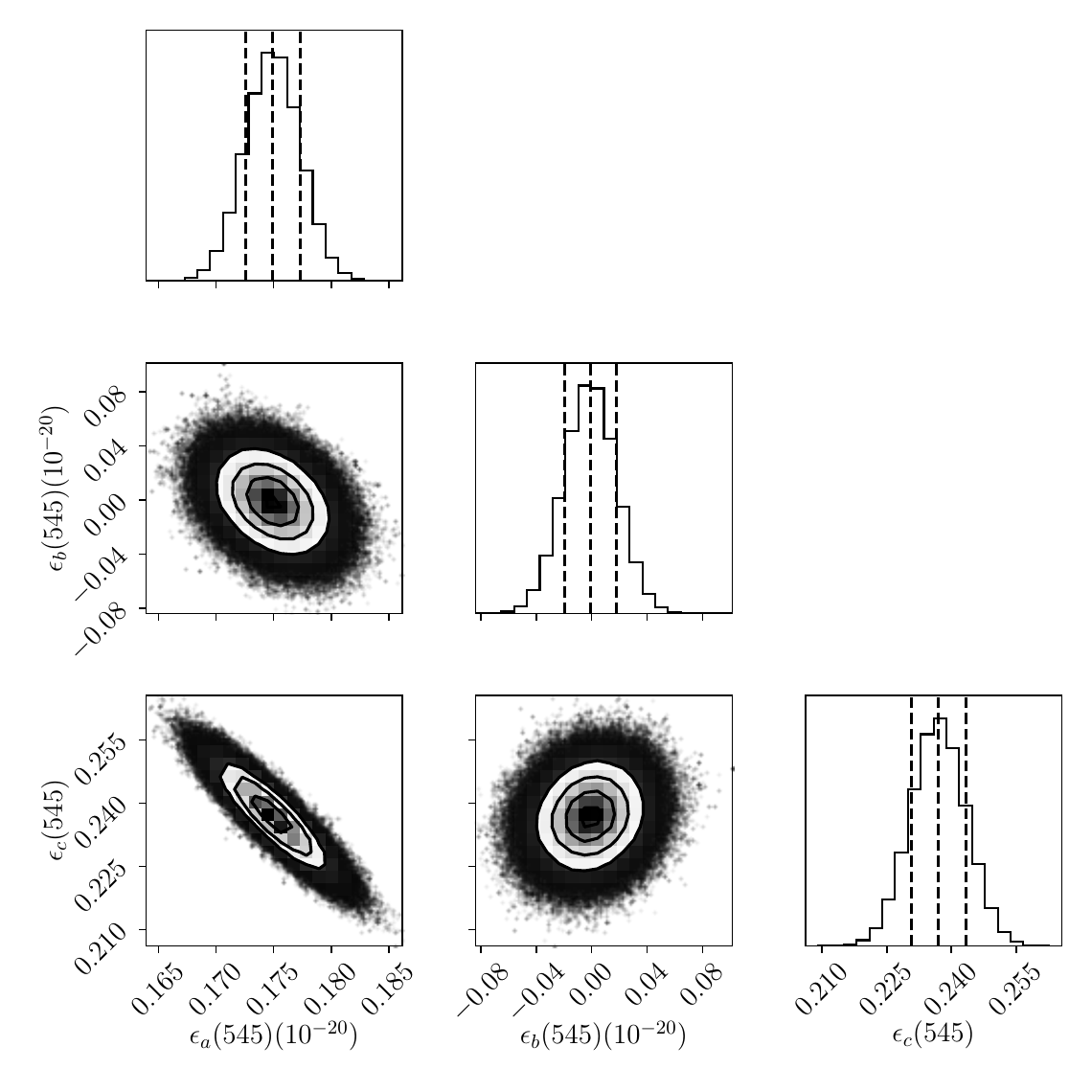}
\end{minipage}
\begin{minipage}{8.5cm}
\includegraphics[width=8.4cm]{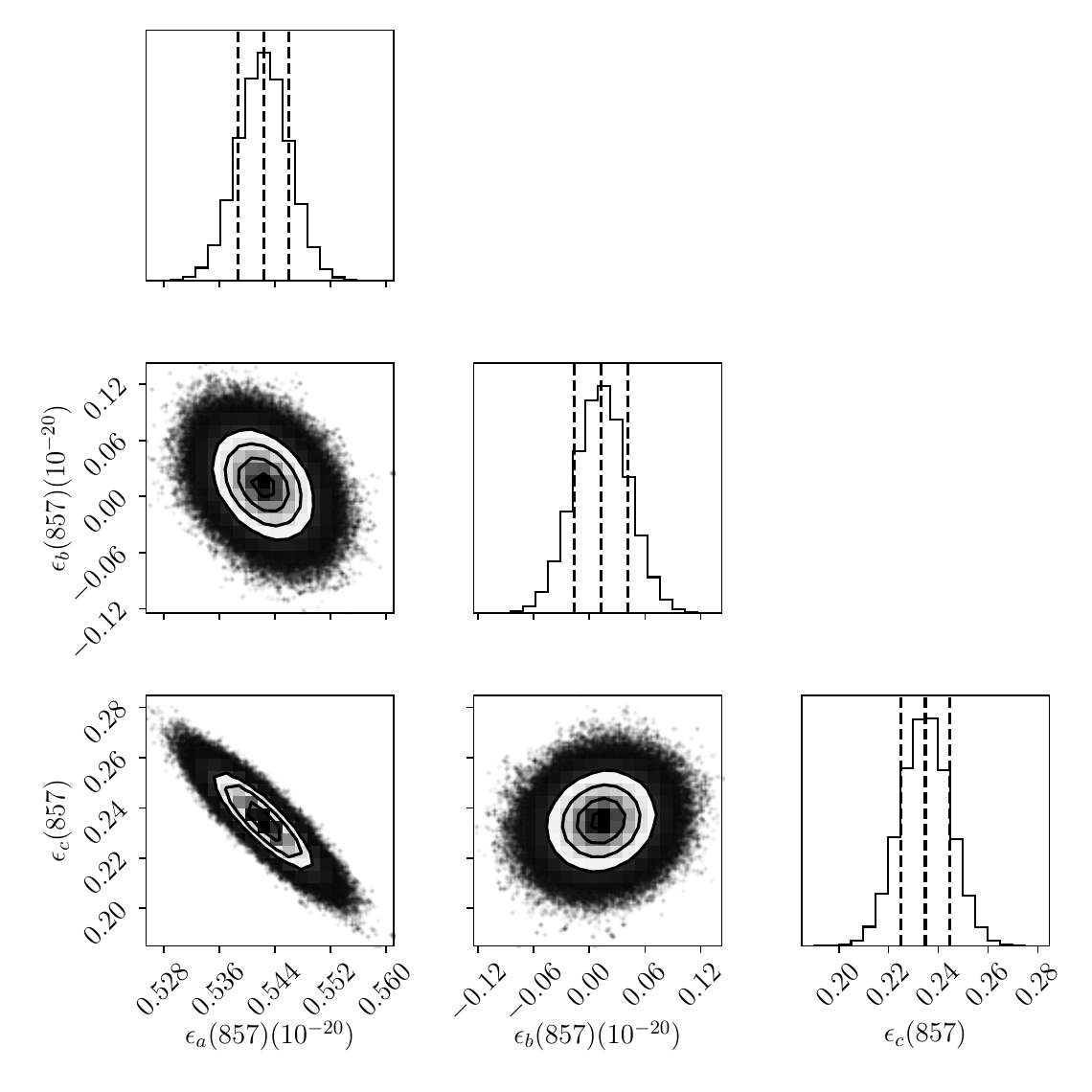}
\end{minipage}
\hfill
\begin{minipage}{8.5cm}
\includegraphics[width=8.4cm]{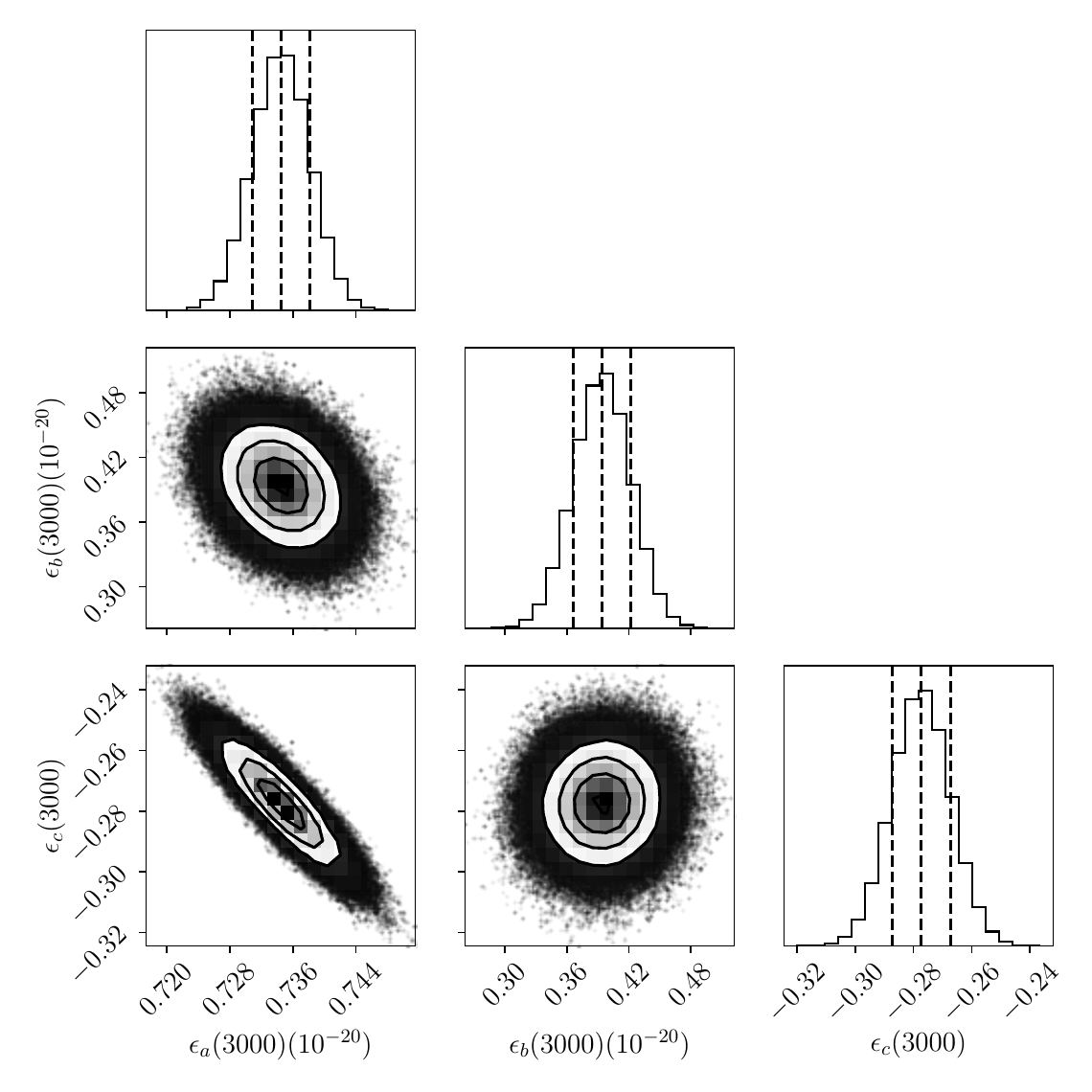}
\end{minipage}

\caption{ \label{fig:planefits} Corner plots showing the posterior distribution results of the MCMC plane fitting of the $\epsilon_a(\nu)$, $\epsilon_a(\nu)$, and $\epsilon_c(\nu)$ parameters at 353 GHz (top left), 545 GHz (top right), 857 GHz (bottom left), and 3000 GHz (bottom right) for the case where $\XHwnm=0.0$ and $\XHwim=1.0$. These plots are the results of fitting the 14 lines-of-sight together where each line-of-sight is given equal weight. The values in Tables 2 and 3 are derived from a distribution of such fits performed using bootstrapping, where in each bootstrap iteration the weights given to each line-of-sight are randomly varied. 
}
\end{figure*}

Parallax measurements either through interferometric measurements or through pulsar timing are the most reliable distance measurements for radio pulsars. Using the pulsar catalogue of \cite{mhth05}\footnote{\url{http://www.atnf.csiro.au/research/pulsar/psrcat/}, version 1.70}, there are seven pulsars that fit our criteria: PSRs B0148--06, B0149--16, B1254--10, B1541+09, J2248--0101, and J2346--0609. Their properties are summarized in Table~\ref{table:1}. 

Radio pulsars in globular clusters also have reliable distance estimates, since the distances to the globular clusters are known. An online compilation by Paulo Freire lists 283 pulsars in 38 globular clusters.\footnote{\url{https://www3.mpifr-bonn.mpg.de/staff/pfreire/GCpsr.html},\\ version dated 2023 April 14} Pulsar dispersion measures within globular clusters show small levels of variation due to the ionized gas within the globular clusters \citep{Freire2001} and due to small angular fluctuations between nearby sightlines from the Galactic $\DM$ \citep{Ransom07}. Since this variation is very small, we consider every globular cluster as a single datum and adopt the mean and the standard deviation of the pulsar $\DM$s within a globular cluster as the representative value and uncertainty of the $\DM$, respectively, for that globular cluster. Distances to globular clusters are taken from \citet{2021MNRAS.505.5957B}. From these sources, we find that seven globular clusters containing pulsars fit our criteria: 47~Tuc, M3, M5, M13, M30, M53, and NGC~362.

In total, 14 sightlines or data points were found to match our criteria. The rest of the paper will only deal with these sightlines and data associated with them. Data relevant to these selected sightlines have been tabulated in Table \ref{table:1}.

\subsection{HI Column Densities}
\label{ssec:hi}

The \HI\ column density is defined as,
\begin{equation}
N_{\rm H\,I} = \int_0^\infty n_{\rm H\,I}(s) \, ds
\end{equation}
where $n_{\rm H\,I}(s)$ is the atomic hydrogen density along the line element $ds$. For high latitudes and restricted to low- and intermediate-velocity \HI\ components, we expect our measured \HI\ column densities to represent the Galactic \HI\ since any extragalactic or HVC contribution has been excluded.

For the selected sightlines, we use the HI4PI all-sky database of Galactic \HI\ \citep{HI4PI}, which is a composite of the GASS III survey \citep{GASS, GASS2, GASS3} and the EBHIS survey \citep{EBHIS}. We use \HI\ column densities integrated for a local standard of rest (LSR) velocity range of $|v| \le 75 \,\, \mathrm{km \, s^{-1}}$. We ignore high-velocity clouds (HVC) in our \HI\ column densities because dust emission is expected to be weak in HVCs. In practice, the only selected sightline that has high-velocity \HI\ contributions is the 47~Tuc sightline which is close to the Small Magellanic Cloud (SMC). In this case, ignoring HVCs simply had the effect of ignoring the SMC, which is ideal since we are concerned with Galactic \HI. For all other sightlines, there are no significant high-velocity components to the overall \HI\ profile. The extracted column densities are listed in Table~\ref{table:1}. The relative error of these column densities is taken to be $3 \%$.

\begin{deluxetable*}{cccc}
\tablecaption{Plane fitting results for $\XHwnm=0.0$ and $\XHwim=1.0$.\label{plane1}}   
\tablehead{\colhead{Frequency (GHz)} & \colhead{$\epsilon_{a}(\nu)$ ($10^{-20} \,\, \mathrm{MJy~cm^2~sr^{-1}}$)} & \colhead{$\epsilon_{b}(\nu)$ ($10^{-20} \,\, \mathrm{MJy~cm^2~sr^{-1}}$)} & \colhead{$\epsilon_c(\nu)$ ($\mathrm{MJy~sr^{-1}}$)}}
\startdata
353 & $0.047{\pm0.008}$ & $0.02{\pm0.03}$ & $0.09{\pm0.03}$ \\ 
545 & $0.17{\pm0.02}$ & $0.05{\pm0.10}$ & $0.2{\pm0.1}$ \\ 
857 & $0.52{\pm0.08}$ & $0.2{\pm0.4}$ & $0.2{\pm0.3}$ \\ 
3000 & $0.74{\pm0.07}$ & $0.1{\pm0.5}$ & $-0.3{\pm0.4}$ \\ 
\enddata
\tablecomments{The $\epsilon_a$ and $\epsilon_b$ parameters correspond to the spectral intensity of dust in the WNM$_\text{\HI}$ and WIM, respectively. Parameter $\epsilon_c$ has no physical meaning in the context of the blackbody fitting, but represents the zero-level offset of each map. The values in this table are the means and standard deviation derived from the bootstrapping with 100 iterations. Thus, they will not exactly correspond to the posterior distributions shown in Figure~1, which are the results of a single iteration.} 
\label{table:num}
\end{deluxetable*}
\begin{deluxetable*}{cccc}
\tablecaption{Plane fitting results for $\XHwnm=0.1$ and $\XHwim=0.9$.\label{plane2}}   
\tablehead{\colhead{Frequency (GHz)} & \colhead{$\epsilon_{a}(\nu)$ ($10^{-20} \,\, \mathrm{MJy~cm^2~sr^{-1}}$)} & \colhead{$\epsilon_{b}(\nu)$ ($10^{-20} \,\, \mathrm{MJy~cm^2~sr^{-1}}$)} & \colhead{$\epsilon_c(\nu)$ ($\mathrm{MJy~sr^{-1}}$)}}
\startdata
353 & $0.044{\pm0.007}$ & $0.02{\pm0.02}$ & $0.09{\pm0.03}$ \\ 
545 & $0.16{\pm0.02}$ & $0.06{\pm0.09}$ & $0.2{\pm0.1}$ \\ 
857 & $0.48{\pm0.07}$ & $0.2{\pm0.3}$ & $0.1{\pm0.3}$ \\ 
3000 & $0.69{\pm0.08}$ & $0.2{\pm0.5}$ & $-0.4{\pm0.4}$ \\ 
\enddata
\tablecomments{The $\epsilon_a$ and $\epsilon_b$ parameters correspond to the spectral intensity of dust in the WNM$_\text{\HI}$ and WIM, respectively. Parameter $\epsilon_c$ has no physical meaning in the context of the blackbody fitting, but represents the zero-level offset of each map.} %
\label{table:num2}
\end{deluxetable*}

\subsection{Dust Emission Intensity}
\label{ssec:dust}

The {\it Planck} mission has produced extremely detailed all-sky maps for dust emission. We use the generalized needlet internal linear combination (GNILC) dust emission maps described by \cite{planck2016-XLVIII} at $353$, $545$, and $857$ GHz. These are the result of a modified black-body fit to a filtered version of the {\it Planck} maps. They have been processed to reduce the cosmic infrared background (CIB) contamination and a zero point correction has been applied. Additionally, we use a fourth $3000$ GHz (100~$\mu$m) dust emission map created from a combination of the IRIS map from \cite{IRIS} and the map from \cite{SFD}. As a whole, we have all-sky dust emission maps for four different frequencies making it possible to measure a 4-point SED of dust for any sightline. 

All the maps are on a HEALPix\footnote{\url{http://healpix.sourceforge.net}} grid \citep{Gorski:2005ku} with $N_{\text{side}}=2048$  (1.72$'/\text{pixel}$). 
The resolution of the GNILC map is variable on the sky, but it is the same at each wavelength for a given position on the sky. The highest resolution is FWHM$=5'$. For our analysis, we use the intensity of the HEALPix pixel in which the pulsar coordinate falls. We do not interpolate more finely between neighbours as the pixel size is much smaller than the beam for all positions. The resulting intensities, $I_{\nu }$, for each sightline are given in Table~\ref{table:1}.

For the total uncertainty, $\delta {I_\nu }$, on these measurements, also given in Table~\ref{table:1} we consider several error terms that we combine in quadrature as:

\begin{align*}
\delta {I_\nu } = \Big[ {{\left( {{\varepsilon _{{\rm{calibration}}}}{\sigma _{{\rm{CMB}}}}} \right)}^2} + {{\left( {{\varepsilon _{{\rm{CIBA}}}}{\sigma _{{\rm{CIBA}}}}} \right)}^2}+ {{\left( {{\varepsilon _{{\rm{noise}}}}{\sigma _{{\rm{noise}}}}} \right)}^2}
\Big] ^{1/2}.
\end{align*}

Here ${{\varepsilon _{{\rm{calibration}}}}{\sigma _{{\rm{CMB}}}}}$ represents the uncertainty on the CMB removal. This term is proportional to the local value of the CMB fluctuations ($\sigma _{{\rm{CMB}}}$) times the calibration uncertainty (${\varepsilon _{{\rm{calibration}}}}$) as a function of frequency. The term, ${{\varepsilon _{{\rm{CIBA}}}}{\sigma _{{\rm{CIBA}}}}}$, is the fraction of the CIB anisotropies left after the GNILC processing, and ${{\varepsilon _{{\rm{noise}}}}{\sigma _{{\rm{noise}}}}}$ is the noise that remains after the GNILC filtering.

From \cite{IRIS} and \cite{planck2014-a08} we estimate $\varepsilon _{{\rm{calibration}}}$ as 0.0078, 0.061, 0.064, and 0.136 at 353, 545, 857, and 3000~GHz respectively. Using \citet[][]{2019ApJ...870..120C}, we estimate  $\varepsilon_{{\rm{noise}}} \approx \varepsilon _{{\rm{CIBA}}} \approx 0.2$ at all frequencies. 

Despite the consideration of these multiple error terms, we expect spatial variation across the sky to dominate the uncertainty in our fits. The 14 lines-of-sight that we use vary widely in position across the sky, and by combining these into a single fit we assume that we can fit an average value across the sky, which likely has a large intrinsic uncertainty. To account for this spatial variation, we use bootstrapping to select different combinations of the lines-of-sight. This is discussed further in Sec.~\ref{ssec:plane} when we describe the fitting.

\section{Analysis}
\label{sec:analysis}

\subsection{Separating the WNM\textsubscript{HI} from the WIM}
\label{sec:wimwnm}

For each of our 14 selected sightlines, $i$, $N_{{\rm H\,I},i}$ provides a measure of the neutral atomic hydrogen gas column along that sightline, and the $\DM_i$ provides a measure of ionized gas column along that sightline. Since the WNM$_\text{\HI}$ and the WIM can both be seen as mixtures of neutral and ionized gas with different ionization fractions, a given measurement of these ionization fractions should allow us to determine how much WNM$_\text{\HI}$ and WIM there is along each sightline.

We use the following set of equations,
\begin{equation}
\label{eqn:DMHI}
\begin{aligned}
&\DM_i &= \XHwnm, \Nwnm + \XHwim, \Nwim + \XHewim \, r \,\Nwim \\
&N_{{\rm H\,I},i} &= \left( 1 - \XHwnm \right) \Nwnm  + \left( 1 - \XHwim \right) \Nwim 
\end{aligned}
\end{equation}
where $\XHwnm$ and $\XHwim$ are the fractional ionizations of hydrogen in the WNM$_\text{\HI}$ and the WIM, respectively, $\XHewim$ is the fractional ionization of singly-ionized helium in the WIM, $r = N_\mathrm{He}/N_\mathrm{H} \approx 0.1$ represents the relative abundance of helium, and $\Nwnm$ and $\Nwim$ are the hydrogen gas column densities of the WNM$_\text{\HI}$ and the WIM, respectively. If the fractional ionizations are known or assumed, this set of equations can be easily solved for $\Nwnm$ and $\Nwim$ for each sightline. Helium in the WNM$_\text{\HI}$ is ignored since it is expected that in the WNM$_\text{\HI}$, helium should be mostly neutral and will not be a significant contributor to DM. 

From \cite{Rey98}, \cite{Mad06}, and \cite{Jenk13}, we obtain reasonable estimates for these fractional ionizations, 
\begin{equation}
\label{eqn:fracs}
\begin{aligned}
\XHwnm &= 0.0 - 0.1 \\
\XHwim &= 0.9 - 1.0 \\
\XHewim &= \left( 0.3 - 0.6 \right) \times \XHwim
\end{aligned}
\end{equation}

As a first estimate, we assume

\begin{equation}
\label{eqn:fracs01}
\begin{aligned}
\XHwnm &= 0.0 \\
\XHwim &= 1.0  \\
\XHewim &= 0.5 \times \XHwim,
\end{aligned}
\end{equation}

which means Eq.~\ref{eqn:DMHI} reduces to

\begin{equation}
\label{eqn:DMHI01}
\begin{aligned}
&\DM_i = 1.05~\Nwim \\
&N_{{\rm H\,I},i} = \Nwnm
\end{aligned}
\end{equation}

We also fit the case where 

\begin{equation}
\label{eqn:fracs0109}
\begin{aligned}
\XHwnm &= 0.1 \\
\XHwim &= 0.9  \\
\XHewim &= 0.5 \times \XHwim,
\end{aligned}
\end{equation}

which gives

\begin{equation}
\label{eqn:DMHI0109}
\begin{aligned}
&\DM_i = 0.1 \Nwnm + 0.95 ~\Nwim \\
&N_{{\rm H\,I},i} = 0.9 \Nwnm + 0.1 ~\Nwim.
\end{aligned}
\end{equation}

We do not include a contribution from $\text{H}_2$ since \cite{planck2011-7.12} showed this to be small (consistent with zero within the uncertainties) for $N_\text{HI} < 4\times10^{20}$~cm$^{-2}$. Only one of the pulsars in Table~\ref{table:1} has $N_\text{HI} > 4\times10^{20}$~cm$^{-2}$, and even for this point, \cite{planck2011-7.12} show that the contribution from $\text{H}_2$ is expected to be $\lesssim20\%\pm15\%$. We are explicitly using the \HI column density to quantify the WNM$_\text{\HI}$ component. This  ignores a possible contribution from the cold neutral medium (CNM).  

\begin{figure*}[!ht]
\centering 
\includegraphics[width=18cm]{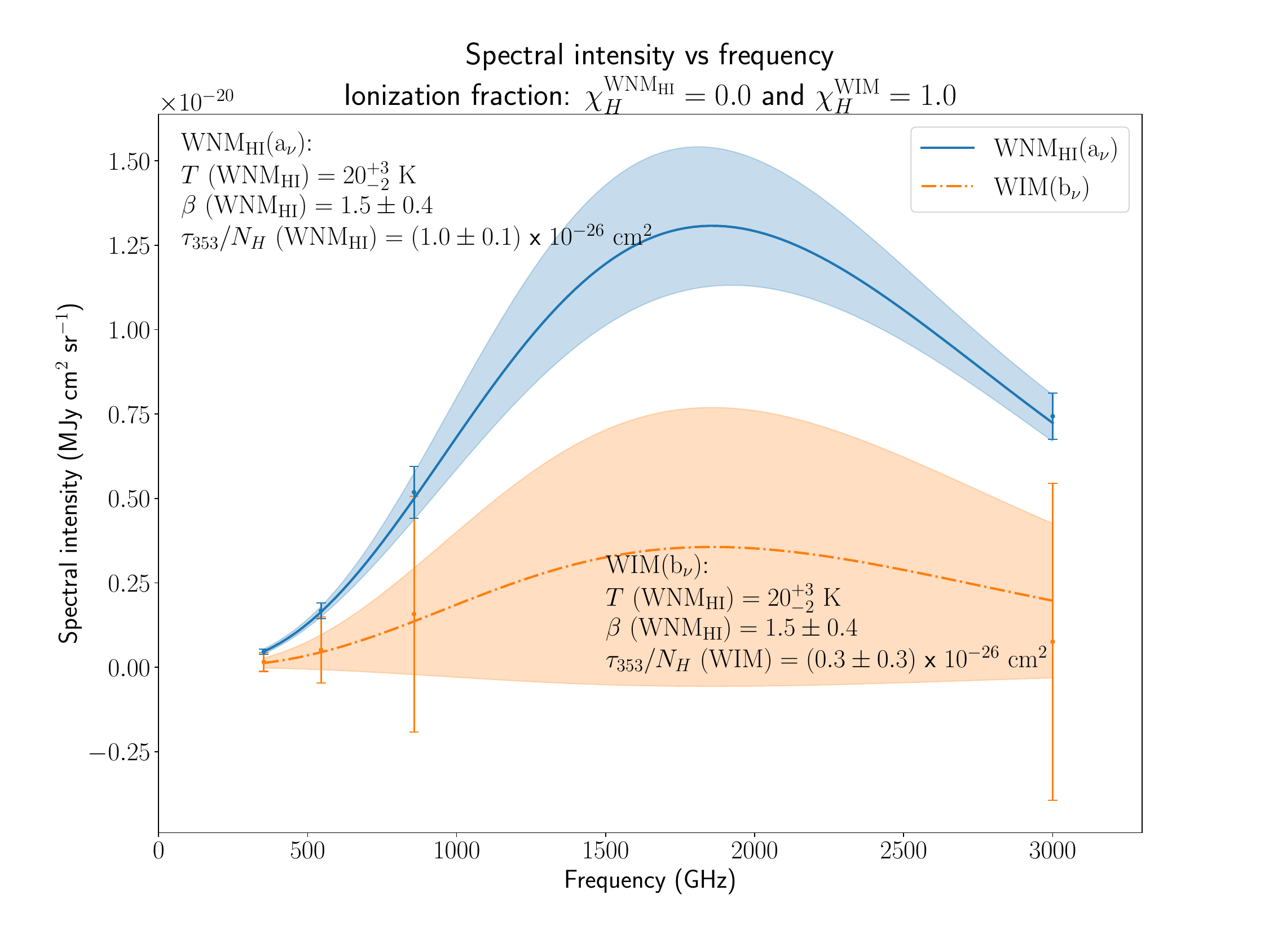}
\caption{ \label{fig:blackbody_fixedTempBeta} Results of the MCMC blackbody fitting. Here we fix $T~\text{(WIM)}=T~\text{(WNM}_\text{\HI}\text{)}$ and $\beta~\text{(WIM)}=\beta~\text{(WNM}_\text{\HI}\text{)}$, in order to fit for $\tau_{\text{353}}/N_\text{H}~\text{(WIM)}$, and $\XHwnm = 0.0 $ and $\XHwim = 1.0$. The lines show the mean of the posterior distribution, and the shaded regions show the 1$\sigma$ confidence regions.}
\end{figure*}

\begin{figure}[!ht]
\centering 
\includegraphics[width=8.3cm]{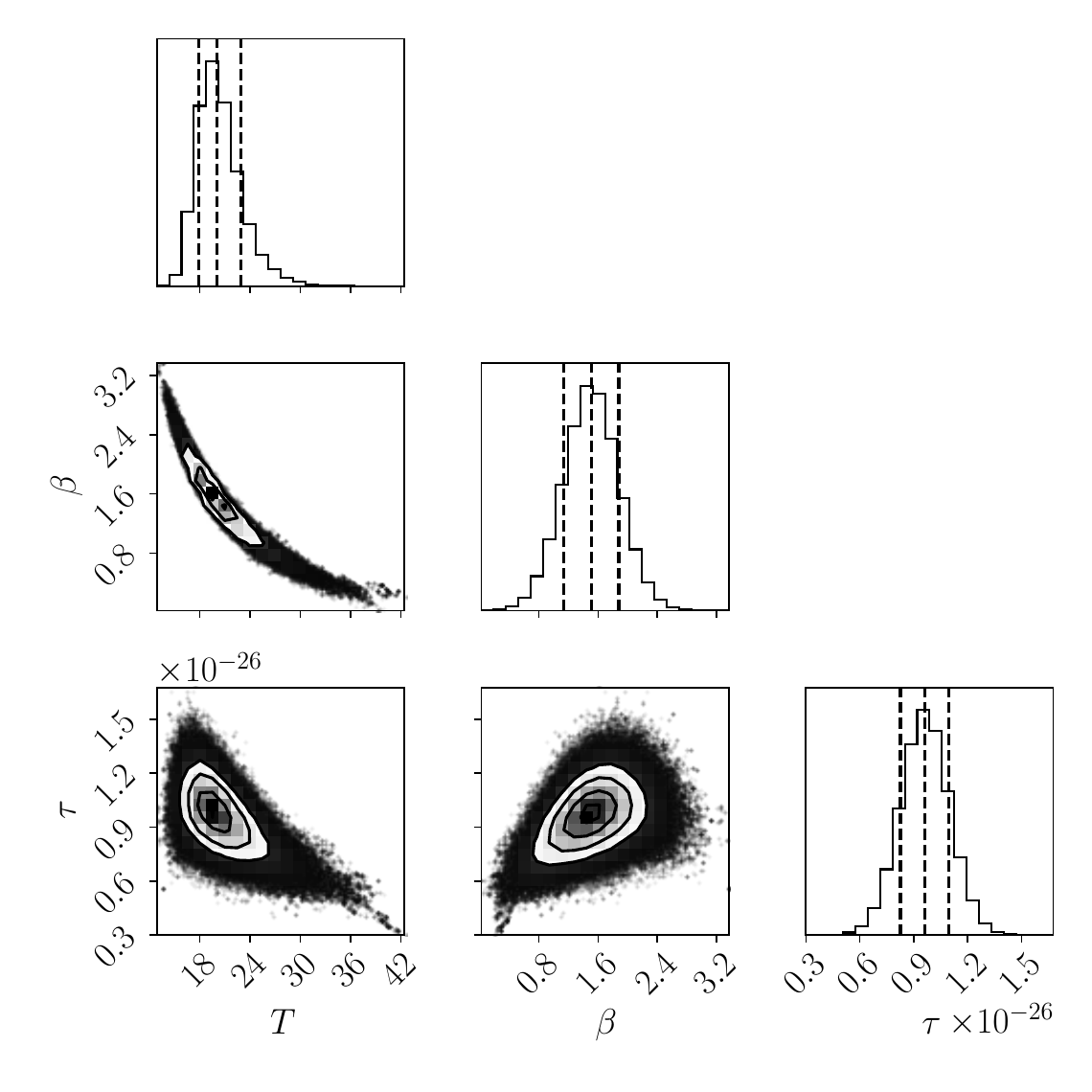}
\caption{ \label{fig:blackbody_cornerplot} Posterior distribution for the MCMC blackbody fitting of the WNM$_\text{\HI}$ parameters.}
\end{figure}

\begin{figure}[!ht]
\centering 
\includegraphics[width=5.5cm]{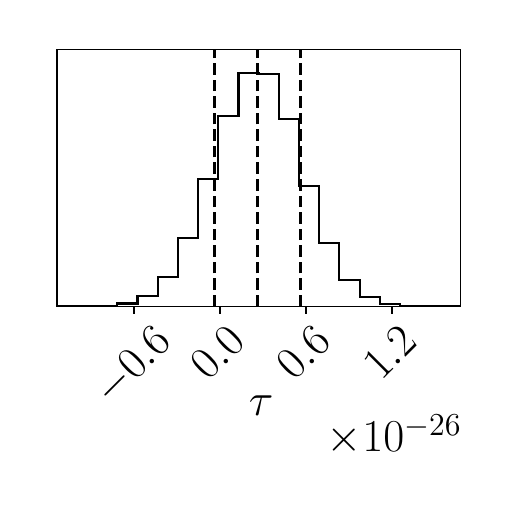}
\caption{ \label{fig:tau_wim_distribution} Posterior distribution for the MCMC fitting of $\tau$(WIM) parameter.}
\end{figure}

\begin{table*}
\centering
\begin{tabular}{|c|c|c|c|c|}
\hline 
 & \multicolumn{2}{c|}{$\XHwnm = 0.0 $, $\XHwim = 1.0$} & \multicolumn{2}{c|}{$\XHwnm = 0.1 $, $\XHwim = 0.9$}\tabularnewline
\hline 
\hline 
 & WNM$_\text{\HI}$ ($\epsilon_{a}(\nu)$) & WIM ($\epsilon_{b}(\nu)$) & WNM$_\text{\HI}$ ($\epsilon_{a}(\nu)$) & WIM ($\epsilon_{b}(\nu)$)\tabularnewline
\hline 
$T$ (K) & \multicolumn{2}{c|}{$20\pm_{2}^{3}$} & \multicolumn{2}{c|}{$20\pm_{2}^{3}$}\tabularnewline
\hline 
$\beta$ & \multicolumn{2}{c|}{$1.5\pm{0.4}$} & \multicolumn{2}{c|}{$1.5\pm{0.4}$}\tabularnewline
\hline 
$\tau_{353}/N_{H}$ ($\times 10^{-26}$~cm$^2$) & $1.0\pm0.1$ & $0.3\pm0.3$ & $0.9\pm{0.1}$  & $0.4\pm0.3$\tabularnewline
\hline 
\end{tabular}

\caption{Means for the fitting of parameters $T$, $\beta$, and $\tau_{\text{353}}/N_\text{H}$ for the WNM$_\text{\HI}$ ($\epsilon_{a}(\nu)$). Also given are the means for fitting $\tau_{\text{353}}/N_\text{H}$ for the WIM ($\epsilon_{b}(\nu)$). The fit for $\tau_{\text{353}}/N_\text{H}$ (WIM) assumes the same values of $T$ and $\beta$ as for the WNM$_\text{\HI}$. We show both cases where $\XHwnm = 0.0 $ and $\XHwim = 1.0$ and where $\XHwnm = 0.1 $ and $\XHwim = 0.9 $.}
\label{table:results_fixed_T_beta}
\end{table*}

\subsection{Plane fitting}
\label{ssec:plane}

Although the {\it Planck} intensities are corrected for the zero point level, the corrections are uncertain. In addition, given that we only have 4 frequencies, we cannot separate the dust SED for the \HI and WIM for each line of sight. To solve this, we instead fit all the pulsars together. For each frequency, $\nu$, and each sightline, $i$ , we have

\begin{align}
I_{i}(\nu) = \epsilon_{a}(\nu) \times \Nwnm + \epsilon_{b}(\nu) \times \Nwim + \epsilon_c(\nu),
\end{align}

which is the same or similar to the forms used in \citet{planck2011-7.12} and \citet{L00}. With our data for the values of $\DM_i$ and $N_{{\rm H\,I},i}$ for each of our 14 sightlines (i.e., values of $i$) in our sample, we can solve for the values of $\Nwim$ and $\Nwnm$ as discussed in Sec.~\ref{sec:wimwnm}. Then we can fit a 2D plane, solving for the parameters $\epsilon_{a}(\nu)$ and $\epsilon_{b}(\nu)$, and $\epsilon_c(\nu)$. 

The parameters $\epsilon_{a}(\nu)$ and $\epsilon_{b}(\nu)$ represent the spectra of the dust (per H atom) in each phase. 

The parameter $\epsilon_c(\nu)$ is a constant that represents the zero level offset. If the zero level correction was well determined, the parameter $\epsilon_c(\nu)$ should not be required, and thus we expect that values for $\epsilon_{c}(\nu)$ should be $<0$, or at least very small. However the accuracy of this correction is uncertain and thus we choose to fit this parameter. We also attempted to fit $\epsilon_{a}(\nu)$ and $\epsilon_{b}(\nu)$ without $\epsilon_c(\nu)$ but the resulting fitted values are unphysical, leading us to conclude that the inclusion of $\epsilon_c(\nu)$ is necessary. We discuss this further in Sec.~\ref{sec:discussion}. 

In order to assess the reasonableness of this approach, we independently check whether the measurements of \HI\ and DM have a correlation with the dust intensity. We use the linregress function from the python scipy stats package \citep{2020SciPy-NMeth} to separately measure the Pearson correlation coefficient of \HI\ with the dust intensity at each of the frequencies, as well as the DM with the dust intensity. We find that the \HI\ has a correlation coefficient $\geq0.96$ at all frequencies. The correlation with DM is weaker, but still measurable having values between $0.36$~--~$0.41$.

We use the python package emcee \citep{2013PASP..125..306F} to perform a Monte-Carlos Markov Chain (MCMC) analysis to find the best fit for $\epsilon_{a}(\nu)$, $\epsilon_{b}(\nu)$, and $\epsilon_{c}(\nu)$. We show the resulting posterior distributions for the 14 lines-of-sight in Fig.~\ref{fig:planefits}. As previously mentioned, the spatial variation on the sky is the dominant source of uncertainty. We estimate the magnitude of this systematic uncertainty by performing our fitting using bootstrapping using 100 draws, each with a sample of 14 random sightlines (with replacement). We find the mean and standard deviation of the distribution. These are summarized in Table~\ref{table:num} for the case where $\XHwnm = 0.0 $ and $\XHwim = 1.0$, and Table~\ref{table:num2} for the case where $\XHwnm = 0.1 $ and $\XHwim = 0.9 $. The former is also plotted as a function of frequency in Fig.~\ref{fig:blackbody_fixedTempBeta}.

\subsection{Blackbody fitting}
\label{ssec:blackbody}

Modeling thermal dust emission has been the subject of many studies \citep{DLee84, DL07, C11}. In the far-infrared to millimetre regime, thermal dust emission is dominated by larger dust grains in thermal equilibrium with the local radiation field. In the optically thin limit, the SED of a uniform population of dust grains can be modeled as a blackbody, which has the following expression,

\begin{equation}
\label{eq:intensity}
I(\nu,i) = \tau_{\nu_0}(i) B_{\nu} (T(i))  \left(\frac{\nu}{\nu_0}\right)^{\beta(i)}
\end{equation}
where 

\begin{equation}
\begin{aligned}
B_\nu(T) = \frac{{2h{\nu ^3}}}{{{c^2}}} \frac { 1 } {e^{h\nu /kT} - 1}.
\end{aligned}
\end{equation}

First we fit the spectrum defined for the measured values for the parameter $\epsilon_{a}(\nu)$ as described in the previous section, for the case where $\XHwnm = 0.0 $ and $\XHwim = 1.0$. $\epsilon_{a}(\nu)$ represents the spectral intensity of the WNM$_\text{\HI}$ component of our total spectrum. For our fitting, we apply bounds on these dust parameters of $2.73$~K$ < T < 60$~K, $-0.5 < \beta < 4.5$, and $0 < \tau_{\text{353}}/N_\text{H} < 1\times 10^{-20}$~cm$^2$. The fit converges and we find the mean of the posterior distributions to be $T~\text{(WNM}_\text{\HI}\text{)}=20^{+3}_{-2}$~K, $\beta~\text{(WNM}_\text{\HI}\text{)} = 1.5\pm{0.4}$, and $\tau_{\text{353}}/N_\text{H}~\text{(WNM}_\text{\HI}\text{)}=(1.0\pm0.1)\times 10^{-26}$~cm$^2$, as shown in Figs.~\ref{fig:blackbody_fixedTempBeta} and \ref{fig:blackbody_cornerplot}.

The parameter $\epsilon_{b}(\nu)$ represents the spectral intensity of the WIM component of our total spectrum. The fitted values of $\epsilon_{b}(\nu)$ are small, with large uncertainties, making this parameter consistent with zero at all frequencies. We are unable to fit for the parameters as a function of $\nu$ using the same bounds as for parameter $\epsilon_{a}(\nu)$ since all points are consistent with zero and the fit does not converge. To address this, we fit only for $\tau_{\text{353}}/N_\text{H}~\text{(WIM)}$ using bounds $-1\times 10^{-20}$~cm$^2 < \tau_{\text{353}}/N_\text{H} < 1\times 10^{-20}$~cm$^2$, while fixing  $T~\text{(WIM)}=T~\text{(WNM}_\text{\HI}\text{)}$ and $\beta~\text{(WIM)}=\beta~\text{(WNM}_\text{\HI}\text{)}$. We find $\tau_{\text{353}}/N_\text{H}~\text{(WIM)}=(0.3\pm0.3)\times 10^{-26}$~cm$^2$. We show the MCMC posterior distribution of $\tau_{\text{353}}/N_\text{H}~\text{(WIM)}$ in Fig.~\ref{fig:tau_wim_distribution}. The fitted blackbody SED is shown in Fig.~\ref{fig:blackbody_fixedTempBeta}. 

By comparing random draws from the posterior distributions of the fits of $\tau_{\text{353}}/N_\text{H}$ for the WNM and the WIM, we show that within this model framework, $\tau_{\text{353}}/N_{H}~\text{(WNM}_\text{\HI}\text{)}$ is at least two times greater than $\tau_{\text{353}}/N_{H}~\text{(WIM)}$ $80\%$ of the time. There is a $21\%$ probability that $\tau_{\text{353}}/N_{H}~\text{(WIM)}\leq0.0$.

We repeat this procedure for the case of $\XHwnm = 0.1 $ and $\XHwim = 0.9$ and find that the fits yield similar values that are consistent within the uncertainties. All of these fitted values are summarized in Table~\ref{table:results_fixed_T_beta}.

\section{Discussion}
\label{sec:discussion}

The dust parameters for the WNM$_\text{\HI}$ component are reasonably well-constrained, but with moderate uncertainties that reflect the expected variation across the sky. The values of $T$ and $\beta$ agree with those found by \cite{planck2013-XVII}, however the value of $\tau_{353} / N_{H}$ is inconsistent within the uncertainty for the case where $\XHwnm = 0.0 $ and $\XHwim = 1.0$. For the case where $\XHwnm = 0.1 $ and $\XHwim = 0.9 $ the values are very close to being in agreement. The discrepancy may be due to the fact that we fit both the WNM$_\text{\HI}$ and WIM components, but \cite{planck2013-XVII} fit only for the WNM$_\text{\HI}$ using the dust-\HI\ correlation.

The parameters for the WIM are less well determined. The uncertainties in our plane fits are consistent with a null result, however when we assume that the values of $T$ and $\beta$ are consistent between the two components, a fit is still possible and yields a value of $\tau_{\text{353}}/N_{H}~\text{(WIM)}$ that is about 1/3 the value for $\tau_{\text{353}}/N_{H}~\text{(WNM}_\text{\HI}\text{)}$.

A particular challenge is determining the uncertainty in the zero level and whether or not to include the parameter $\epsilon_c(\nu)$ in our fitting. When we attempted to fit $\epsilon_{a}(\nu)$ and $\epsilon_{b}(\nu)$ without $\epsilon_c(\nu)$, we found unphysical values of $\epsilon_{b}(\nu)$. For example, the fit then yields $\epsilon_{b}(3000) = (-0.29\pm0.22)\times10^{-20}~\mathrm{MJy~cm^2~sr^{-1}}$, which is significantly less than zero. Additionally, in this case, $\epsilon_{b}(545)$ and $\epsilon_{b}(353)$ are both greater than $\epsilon_{a}(\nu)$, which is unexpected. However, we also expect that values for $\epsilon_{c}(\nu)$ should be $<0$, or at least very small. The fitted values at 545 and 857~GHz, shown in  Tables~\ref{table:num} and \ref{table:num2}, are unexpectedly (and unrealistically) large. The difficulties we have with fitting (or not fitting) this offset might at least in part be due to the small dynamic range in the values of DM. This may be a result of the relative low density and isotropy in the WIM of the Galactic halo, but it nevertheless makes fitting challenging. Having more data points may resolve this difficulty.

From this analysis, we can infer that either there is no dust present, which contradicts previous results, or that $\tau_{\text{353}}/N_\text{H}$ is measurably smaller in the WIM compared to the WNM$_\text{\HI}$. Our assumption that the WNM$_\text{\HI}$ and WIM have the same $T$ and $\beta$ implies that these components would have a similar chemical composition, dust grain shape and size, and thermal properties. The significantly lower dust opacity in the WIM compared to the WNM$_\text{\HI}$, suggests that there is less dust in the WIM compared to the WNM$_\text{\HI}$. 

A helpful addition to the fitting algorithm could be adding constraints of dust abundances from 3D dust reddening maps. Currently, the map with the farthest distance resolution is Bayestar \cite{Green2019}. This was not used in this study because out of the 14 line of sights we considered, 3 were in directions not covered by the map, and 5 were beyond the reliable distance of the dust map. In the future, as more data becomes available, this additional information can be included in the algorithm.

\section{Conclusion}
\label{sec:conclusion}

Using pulsar dispersion measures, \HI\ column densities, and thermal dust emission data from the {\it Planck} mission, we attempt a component separation on dust emission to characterize dust in the WIM and the WNM$_\text{\HI}$. From our fitting, we find $T~\text{(WNM}_\text{\HI}\text{)}=20^{+3}_{-2}$~K, $\beta~\text{(WNM}_\text{\HI}\text{)} = 1.5\pm{0.4}$, and $\tau_{\text{353}}/N_{H}~\text{(WNM}_\text{\HI}\text{)}=(1.0\pm0.1)\times 10^{-26}$~cm$^2$. For the WIM, we find that $\tau_{\text{353}}/N_{H}~\text{(WIM)}=(0.3\pm0.3)\times 10^{-26}$, which is about three times smaller than $\tau_{\text{353}}/N_{H}~\text{(WNM}_\text{\HI}\text{)}$. We are 80\% confident that $\tau_{\text{353}}/N_{H}~\text{(WIM)}$ is at least two times smaller than $\tau_{\text{353}}/N_{H}~\text{(WNM}_\text{\HI}\text{)}$, but we emphasize that this result comes from a simple model with an inherent assumption that we can compute an average across value of $\tau_{\text{353}}/N_{H}$ using very widely separated lines-of-sight. Using this model, we find that the WIM has a dust opacity much less than that of the dust in the WNM$_\text{\HI}$, implying that there is less dust in the WIM than in the WNM$_\text{\HI}$. 

The large uncertainty in this work results from this being a difficult measurement that uses relatively few lines-of-sight and assumes that an average fit across the whole sky is possible. This issue may be resolved in the future with many more lines-of-sight included in the fitting. These additional lines-of-sight may come from newly discovered halo pulsars. More and better pulsar distances should be released over time from VLBI parallax measurements. It might be valuable to study the dust in the WIM of the Large Magellanic Cloud, given the wealth of pulsars discovered within it.

Fast Radio Bursts are also an exciting new probe of Galactic DM \citep{2023ApJ...946...58C}. In the near future, all sky DM maps such as the one presented by \citet{2023arXiv230412350H}, may allow our method to be used across the whole sky. Future studies should also measure dust emission at frequencies between 1500 and 2500 GHz to better constrain the fitting. 
\\
\\
%\begin{acknowledgements}
We thank Chris Howk, Alex Hill, Bob Benjamin, Edith Falgarone, Katia Ferri\`{e}re, Patrick Hennebelle, Josh Peek, and Tessa Vernstrom for vaulable discussions that improved this manuscript. The Dunlap Institute is funded through an endowment established by the David Dunlap family and the University of Toronto. B.M.G. acknowledges the support of the Natural Sciences and Engineering Research Council of Canada (NSERC) through grants RGPIN-2015-05948 and RGPIN-2022-03163, and of the Canada Research Chairs program, and thanks the \'Ecole Normale Sup\'erieure for hosting the visit during which this work began. This work benefited from discussions during the Interstellar Institue II2, {\it The Milky-Way in the age of Gaia} (October 2018), the  Interstellar Institue II5, {\it With Two Eyes} (July 2022), and the program {\it Towards a Comprehensive Model of the Galactic Magnetic Field} hosted at Nordita (April 2023) by the IMAGINE consortium, which was partly supported by NordForsk and the Royal Astronomical Society. 
%\end{acknowledgements}

\bibliographystyle{yahapj}
\bibliography{references}

\begin{thebibliography}{}
\providecommand\natexlab[1]{#1}
\providecommand\JournalTitle[1]{#1}

\bibitem[{{Arendt} {et~al.}(1998){Arendt}, {Odegard}, {Weiland}, {Sodroski},
  {Hauser}, {Dwek}, {Kelsall}, {Moseley}, {Silverberg}, {Leisawitz},
  {Mitchell}, {Reach}, \& {Wright}}]{1998ApJ...508...74A}
{Arendt}, R.~G., {Odegard}, N., {Weiland}, J.~L., {et~al.} 1998,
  \href{http://dx.doi.org/10.1086/306381}{\JournalTitle{\apj}, 508, 74}

\bibitem[{{Balakrishnan} {et~al.}(2023){Balakrishnan}, {Freire}, {Ransom},
  {Ridolfi}, {Barr}, {Chen}, {Krishnan}, {Champion}, {Kramer}, {Gautam},
  {Padmanabh}, {Men}, {Abbate}, {Stappers}, {Stairs}, {Keane}, \&
  {Possenti}}]{2023ApJ...942L..35B}
{Balakrishnan}, V., {Freire}, P. C.~C., {Ransom}, S.~M., {et~al.} 2023,
  \href{http://dx.doi.org/10.3847/2041-8213/acae99}{\JournalTitle{\apjl}, 942,
  L35}

\bibitem[{{Baumgardt} \& {Vasiliev}(2021)}]{2021MNRAS.505.5957B}
{Baumgardt}, H., \& {Vasiliev}, E. 2021,
  \href{http://dx.doi.org/10.1093/mnras/stab1474}{\JournalTitle{\mnras}, 505,
  5957}

\bibitem[{{Bilous} {et~al.}(2016){Bilous}, {Kondratiev}, {Kramer}, {Keane},
  {Hessels}, {Stappers}, {Malofeev}, {Sobey}, {Breton}, {Cooper}, {Falcke},
  {Karastergiou}, {Michilli}, {Os{\l}owski}, {Sanidas}, {ter Veen}, {van
  Leeuwen}, {Verbiest}, {Weltevrede}, {Zarka}, {Grie{\ss}meier}, {Serylak},
  {Bell}, {Broderick}, {Eisl{\"o}ffel}, {Markoff}, \& {Rowlinson}}]{Bilous2016}
{Bilous}, A.~V., {Kondratiev}, V.~I., {Kramer}, M., {et~al.} 2016,
  \href{http://dx.doi.org/10.1051/0004-6361/201527702}{\JournalTitle{\aap},
  591, A134}

\bibitem[{{Boulanger} {et~al.}(1996){Boulanger}, {Abergel}, {Bernard},
  {Burton}, {Desert}, {Hartmann}, {Lagache}, \& {Puget}}]{Bou96}
{Boulanger}, F., {Abergel}, A., {Bernard}, J.-P., {et~al.} 1996,
  \JournalTitle{\aap}, 312, 256

\bibitem[{{Boulanger} \& {Perault}(1988)}]{1988ApJ...330..964B}
{Boulanger}, F., \& {Perault}, M. 1988,
  \href{http://dx.doi.org/10.1086/166526}{\JournalTitle{\apj}, 330, 964}

\bibitem[{{Chiang} \& {M{\'e}nard}(2019)}]{2019ApJ...870..120C}
{Chiang}, Y.-K., \& {M{\'e}nard}, B. 2019,
  \href{http://dx.doi.org/10.3847/1538-4357/aaf4f6}{\JournalTitle{\apj}, 870,
  120}

\bibitem[{{Clark} {et~al.}(2019){Clark}, {Peek}, \&
  {Miville-Desch{\^e}nes}}]{2019ApJ...874..171C}
{Clark}, S.~E., {Peek}, J.~E.~G., \& {Miville-Desch{\^e}nes}, M.~A. 2019,
  \href{http://dx.doi.org/10.3847/1538-4357/ab0b3b}{\JournalTitle{\apj}, 874,
  171}

\bibitem[{{Compi{\`e}gne} {et~al.}(2011){Compi{\`e}gne}, {Verstraete}, {Jones},
  {Bernard}, {Boulanger}, {Flagey}, {Le Bourlot}, {Paradis}, \& {Ysard}}]{C11}
{Compi{\`e}gne}, M., {Verstraete}, L., {Jones}, A., {et~al.} 2011,
  \href{http://dx.doi.org/10.1051/0004-6361/201015292}{\JournalTitle{\aap},
  525, A103}

\bibitem[{{Cook} {et~al.}(2023){Cook}, {Bhardwaj}, {Gaensler}, {Scholz},
  {Eadie}, {Hill}, {Kaspi}, {Masui}, {Curtin}, {Dong}, {Fonseca},
  {Herrera-Martin}, {Kaczmarek}, {Lanman}, {Lazda}, {Leung}, {Meyers},
  {Michilli}, {Pandhi}, {Pearlman}, {Pleunis}, {Ransom}, {Rahman}, {Sand},
  {Shin}, {Smith}, {Stairs}, \& {Stenning}}]{2023ApJ...946...58C}
{Cook}, A.~M., {Bhardwaj}, M., {Gaensler}, B.~M., {et~al.} 2023,
  \href{http://dx.doi.org/10.3847/1538-4357/acbbd0}{\JournalTitle{\apj}, 946,
  58}

\bibitem[{{Deller} {et~al.}(2019){Deller}, {Goss}, {Brisken}, {Chatterjee},
  {Cordes}, {Janssen}, {Kovalev}, {Lazio}, {Petrov}, {Stappers}, \&
  {Lyne}}]{2019ApJ...875..100D}
{Deller}, A.~T., {Goss}, W.~M., {Brisken}, W.~F., {et~al.} 2019,
  \href{http://dx.doi.org/10.3847/1538-4357/ab11c7}{\JournalTitle{\apj}, 875,
  100}

\bibitem[{Draine(2011)}]{draine2011c}
Draine, B.~T. 2011, Physics of the {{Interstellar}} and {{Intergalactic
  Medium}} (Princeton University Press)

\bibitem[{{Draine} \& {Lee}(1984)}]{DLee84}
{Draine}, B.~T., \& {Lee}, H.~M. 1984,
  \href{http://dx.doi.org/10.1086/162480}{\JournalTitle{\apj}, 285, 89}

\bibitem[{{Draine} \& {Li}(2007)}]{DL07}
{Draine}, B.~T., \& {Li}, A. 2007,
  \href{http://dx.doi.org/10.1086/511055}{\JournalTitle{\apj}, 657, 810}

\bibitem[{{Ferri{\`e}re}(2001)}]{Ferr}
{Ferri{\`e}re}, K.~M. 2001,
  \href{http://dx.doi.org/10.1103/RevModPhys.73.1031}{\JournalTitle{Reviews of
  Modern Physics}, 73, 1031}

\bibitem[{{Foreman-Mackey} {et~al.}(2013){Foreman-Mackey}, {Hogg}, {Lang}, \&
  {Goodman}}]{2013PASP..125..306F}
{Foreman-Mackey}, D., {Hogg}, D.~W., {Lang}, D., \& {Goodman}, J. 2013,
  \href{http://dx.doi.org/10.1086/670067}{\JournalTitle{\pasp}, 125, 306}

\bibitem[{{Freire} {et~al.}(2001){Freire}, {Kramer}, {Lyne}, {Camilo},
  {Manchester}, \& {D'Amico}}]{Freire2001}
{Freire}, P.~C., {Kramer}, M., {Lyne}, A.~G., {et~al.} 2001,
  \href{http://dx.doi.org/10.1086/323248}{\JournalTitle{\apjl}, 557, L105}

\bibitem[{{Gaensler} {et~al.}(2008){Gaensler}, {Madsen}, {Chatterjee}, \&
  {Mao}}]{G08}
{Gaensler}, B.~M., {Madsen}, G.~J., {Chatterjee}, S., \& {Mao}, S.~A. 2008,
  \href{http://dx.doi.org/10.1071/AS08004}{\JournalTitle{\pasa}, 25, 184}

\bibitem[{Gorski {et~al.}(2005)Gorski, Hivon, Banday, Wandelt, Hansen,
  Reinecke, \& Bartelmann}]{Gorski:2005ku}
Gorski, K.~M., Hivon, E., Banday, A.~J., {et~al.} 2005, \JournalTitle{\apj},
  622, 759

\bibitem[{Green {et~al.}(2019)Green, Schlafly, Zucker, Speagle, \&
  Finkbeiner}]{Green2019}
Green, G.~M., Schlafly, E., Zucker, C., Speagle, J.~S., \& Finkbeiner, D. 2019,
  \href{http://dx.doi.org/10.3847/1538-4357/ab5362}{\JournalTitle{The
  Astrophysical Journal}, 887, 93}

\bibitem[{{Haffner} {et~al.}(2009){Haffner}, {Dettmar}, {Beckman}, {Wood},
  {Slavin}, {Giammanco}, {Madsen}, {Zurita}, \& {Reynolds}}]{H09}
{Haffner}, L.~M., {Dettmar}, R.-J., {Beckman}, J.~E., {et~al.} 2009,
  \href{http://dx.doi.org/10.1103/RevModPhys.81.969}{\JournalTitle{Reviews of
  Modern Physics}, 81, 969}

\bibitem[{{Heiles} {et~al.}(1999){Heiles}, {Haffner}, \&
  {Reynolds}}]{1999ASPC..168..211H}
{Heiles}, C., {Haffner}, L.~M., \& {Reynolds}, R.~J. 1999, in Astronomical
  Society of the Pacific Conference Series, Vol. 168, New Perspectives on the
  Interstellar Medium, ed. A.~R. {Taylor}, T.~L. {Landecker}, \& G.~{Joncas},
  211

\bibitem[{{HI4PI Collaboration} {et~al.}(2016){HI4PI Collaboration}, {Ben
  Bekhti}, {Fl{\"o}er}, {Keller}, {Kerp}, {Lenz}, {Winkel}, {Bailin},
  {Calabretta}, {Dedes}, {Ford}, {Gibson}, {Haud}, {Janowiecki}, {Kalberla},
  {Lockman}, {McClure-Griffiths}, {Murphy}, {Nakanishi}, {Pisano}, \&
  {Staveley-Smith}}]{HI4PI}
{HI4PI Collaboration}, {Ben Bekhti}, N., {Fl{\"o}er}, L., {et~al.} 2016,
  \href{http://dx.doi.org/10.1051/0004-6361/201629178}{\JournalTitle{\aap},
  594, A116}

\bibitem[{{Hobbs} {et~al.}(2004){Hobbs}, {Lyne}, {Kramer}, {Martin}, \&
  {Jordan}}]{2004MNRAS.353.1311H}
{Hobbs}, G., {Lyne}, A.~G., {Kramer}, M., {Martin}, C.~E., \& {Jordan}, C.
  2004,
  \href{http://dx.doi.org/10.1111/j.1365-2966.2004.08157.x}{\JournalTitle{\mnras},
  353, 1311}

\bibitem[{{Howk} \& {Consiglio}(2012)}]{howk2012}
{Howk}, J.~C., \& {Consiglio}, S.~M. 2012,
  \href{http://dx.doi.org/10.1088/0004-637X/759/2/97}{\JournalTitle{\apj}, 759,
  97}

\bibitem[{{Howk} {et~al.}(2003){Howk}, {Sembach}, \& {Savage}}]{Howk03}
{Howk}, J.~C., {Sembach}, K.~R., \& {Savage}, B.~D. 2003,
  \href{http://dx.doi.org/10.1086/346262}{\JournalTitle{\apj}, 586, 249}

\bibitem[{{Howk} {et~al.}(2006){Howk}, {Sembach}, \&
  {Savage}}]{2006ApJ...637..333H}
---. 2006, \href{http://dx.doi.org/10.1086/497352}{\JournalTitle{\apj}, 637,
  333}

\bibitem[{{Hutschenreuter} {et~al.}(2023){Hutschenreuter}, {Haverkorn},
  {Frank}, {Raycheva}, \& {En{\ss}lin}}]{2023arXiv230412350H}
{Hutschenreuter}, S., {Haverkorn}, M., {Frank}, P., {Raycheva}, N.~C., \&
  {En{\ss}lin}, T.~A. 2023,
  \href{http://dx.doi.org/10.48550/arXiv.2304.12350}{\JournalTitle{arXiv
  e-prints}, arXiv:2304.12350}

\bibitem[{{Jeli{\'c}} {et~al.}(2018){Jeli{\'c}}, {Prelogovi{\'c}}, {Haverkorn},
  {Remeijn}, \& {Klind{\v{z}}i{\'c}}}]{2018A&A...615L...3J}
{Jeli{\'c}}, V., {Prelogovi{\'c}}, D., {Haverkorn}, M., {Remeijn}, J., \&
  {Klind{\v{z}}i{\'c}}, D. 2018,
  \href{http://dx.doi.org/10.1051/0004-6361/201833291}{\JournalTitle{\aap},
  615, L3}

\bibitem[{{Jenkins}(2013)}]{Jenk13}
{Jenkins}, E.~B. 2013,
  \href{http://dx.doi.org/10.1088/0004-637X/764/1/25}{\JournalTitle{\apj}, 764,
  25}

\bibitem[{{Kalberla} \& {Haud}(2015)}]{GASS2}
{Kalberla}, P.~M.~W., \& {Haud}, U. 2015,
  \href{http://dx.doi.org/10.1051/0004-6361/201525859}{\JournalTitle{\aap},
  578, A78}

\bibitem[{{Kalberla} {et~al.}(2010){Kalberla}, {McClure-Griffiths}, {Pisano},
  {Calabretta}, {Ford}, {Lockman}, {Staveley-Smith}, {Kerp}, {Winkel},
  {Murphy}, \& {Newton-McGee}}]{GASS3}
{Kalberla}, P.~M.~W., {McClure-Griffiths}, N.~M., {Pisano}, D.~J., {et~al.}
  2010,
  \href{http://dx.doi.org/10.1051/0004-6361/200913979}{\JournalTitle{\aap},
  521, A17}

\bibitem[{{Lagache} {et~al.}(1999){Lagache}, {Abergel}, {Boulanger},
  {D{\'e}sert}, \& {Puget}}]{L99}
{Lagache}, G., {Abergel}, A., {Boulanger}, F., {D{\'e}sert}, F.~X., \& {Puget},
  J.-L. 1999, \JournalTitle{\aap}, 344, 322

\bibitem[{{Lagache} {et~al.}(2000){Lagache}, {Haffner}, {Reynolds}, \&
  {Tufte}}]{L00}
{Lagache}, G., {Haffner}, L.~M., {Reynolds}, R.~J., \& {Tufte}, S.~L. 2000,
  \JournalTitle{\aap}, 354, 247

\bibitem[{{Madsen} {et~al.}(2006){Madsen}, {Reynolds}, \& {Haffner}}]{Mad06}
{Madsen}, G.~J., {Reynolds}, R.~J., \& {Haffner}, L.~M. 2006,
  \href{http://dx.doi.org/10.1086/508441}{\JournalTitle{\apj}, 652, 401}

\bibitem[{{Manchester} {et~al.}(2005){Manchester}, {Hobbs}, {Teoh}, \&
  {Hobbs}}]{mhth05}
{Manchester}, R.~N., {Hobbs}, G.~B., {Teoh}, A., \& {Hobbs}, M. 2005,
  \href{http://dx.doi.org/10.1086/428488}{\JournalTitle{\aj}, 129, 1993}

\bibitem[{{McClure-Griffiths} {et~al.}(2023){McClure-Griffiths},
  Stanimirovi{\'c}, \& Rybarczyk}]{mcclure-griffiths2023}
{McClure-Griffiths}, N.~M., Stanimirovi{\'c}, S., \& Rybarczyk, D.~R. 2023,
  \href{http://dx.doi.org/10.1146/annurev-astro-052920-104851}{\JournalTitle{Annual
  Review of Astronomy and Astrophysics}, 61, 19}

\bibitem[{{McClure-Griffiths} {et~al.}(2009){McClure-Griffiths}, {Pisano},
  {Calabretta}, {Ford}, {Lockman}, {Staveley-Smith}, {Kalberla}, {Bailin},
  {Dedes}, {Janowiecki}, {Gibson}, {Murphy}, {Nakanishi}, \&
  {Newton-McGee}}]{GASS}
{McClure-Griffiths}, N.~M., {Pisano}, D.~J., {Calabretta}, M.~R., {et~al.}
  2009,
  \href{http://dx.doi.org/10.1088/0067-0049/181/2/398}{\JournalTitle{\apjs},
  181, 398}

\bibitem[{{Miville-Desch{\^e}nes} \& {Lagache}(2005)}]{IRIS}
{Miville-Desch{\^e}nes}, M.-A., \& {Lagache}, G. 2005,
  \href{http://dx.doi.org/10.1086/427938}{\JournalTitle{\apjs}, 157, 302}

\bibitem[{{Ocker} {et~al.}(2020){Ocker}, {Cordes}, \&
  {Chatterjee}}]{2020ApJ...897..124O}
{Ocker}, S.~K., {Cordes}, J.~M., \& {Chatterjee}, S. 2020,
  \href{http://dx.doi.org/10.3847/1538-4357/ab98f9}{\JournalTitle{\apj}, 897,
  124}

\bibitem[{{Pan} {et~al.}(2021){Pan}, {Qian}, {Ma}, {Liu}, {Wang}, {Luo}, {Yan},
  {Ransom}, {Lorimer}, {Li}, \& {Jiang}}]{2021ApJ...915L..28P}
{Pan}, Z., {Qian}, L., {Ma}, X., {et~al.} 2021,
  \href{http://dx.doi.org/10.3847/2041-8213/ac0bbd}{\JournalTitle{\apjl}, 915,
  L28}

\bibitem[{{Planck Collaboration Int. XLVIII}(2016)}]{planck2016-XLVIII}
{Planck Collaboration Int. XLVIII}. 2016,
  \href{http://dx.doi.org/10.1051/0004-6361/201629022}{\JournalTitle{\aap},
  596, A109}

\bibitem[{{Planck Collaboration Int. XVII}(2014)}]{planck2013-XVII}
{Planck Collaboration Int. XVII}. 2014,
  \href{http://dx.doi.org/10.1051/0004-6361/201323270}{\JournalTitle{\aap},
  566, A55}

\bibitem[{{Planck Collaboration VII}(2016)}]{planck2014-a08}
{Planck Collaboration VII}. 2016,
  \href{http://dx.doi.org/10.1051/0004-6361/201525844}{\JournalTitle{\aap},
  594, A7}

\bibitem[{{Planck Collaboration XI}(2014)}]{planck2013-p06b}
{Planck Collaboration XI}. 2014,
  \href{http://dx.doi.org/10.1051/0004-6361/201323195}{\JournalTitle{\aap},
  571, A11}

\bibitem[{{Planck Collaboration XXIV}(2011)}]{planck2011-7.12}
{Planck Collaboration XXIV}. 2011,
  \href{http://dx.doi.org/10.1051/0004-6361/201116485}{\JournalTitle{\aap},
  536, A24}

\bibitem[{{Ransom}(2007)}]{Ransom07}
{Ransom}, S.~M. 2007,
  \href{http://dx.doi.org/10.48550/arXiv.astro-ph/0611672}{in Astronomical
  Society of the Pacific Conference Series, Vol. 365, SINS - Small Ionized and
  Neutral Structures in the Diffuse Interstellar Medium, ed. M.~{Haverkorn} \&
  W.~M. {Goss}}, 265

\bibitem[{{Reardon} {et~al.}(2021){Reardon}, {Shannon}, {Cameron}, {Goncharov},
  {Hobbs}, {Middleton}, {Shamohammadi}, {Thyagarajan}, {Bailes}, {Bhat}, {Dai},
  {Kerr}, {Manchester}, {Russell}, {Spiewak}, {Wang}, \&
  {Zhu}}]{2021MNRAS.507.2137R}
{Reardon}, D.~J., {Shannon}, R.~M., {Cameron}, A.~D., {et~al.} 2021,
  \href{http://dx.doi.org/10.1093/mnras/stab1990}{\JournalTitle{\mnras}, 507,
  2137}

\bibitem[{{Reynolds} {et~al.}(1998){Reynolds}, {Hausen}, {Tufte}, \&
  {Haffner}}]{Rey98}
{Reynolds}, R.~J., {Hausen}, N.~R., {Tufte}, S.~L., \& {Haffner}, L.~M. 1998,
  \href{http://dx.doi.org/10.1086/311154}{\JournalTitle{\apjl}, 494, L99}

\bibitem[{{Ridolfi} {et~al.}(2021){Ridolfi}, {Gautam}, {Freire}, {Ransom},
  {Buchner}, {Possenti}, {Venkatraman Krishnan}, {Bailes}, {Kramer},
  {Stappers}, {Abbate}, {Barr}, {Burgay}, {Camilo}, {Corongiu}, {Jameson},
  {Padmanabh}, {Vleeschower}, {Champion}, {Chen}, {Geyer}, {Karastergiou},
  {Karuppusamy}, {Parthasarathy}, {Reardon}, {Serylak}, {Shannon}, \&
  {Spiewak}}]{2021MNRAS.504.1407R}
{Ridolfi}, A., {Gautam}, T., {Freire}, P.~C.~C., {et~al.} 2021,
  \href{http://dx.doi.org/10.1093/mnras/stab790}{\JournalTitle{\mnras}, 504,
  1407}

\bibitem[{Saintonge \& Catinella(2022)}]{saintonge2022}
Saintonge, A., \& Catinella, B. 2022,
  \href{http://dx.doi.org/10.1146/annurev-astro-021022-043545}{\JournalTitle{Annual
  Review of Astronomy and Astrophysics}, 60, 319}

\bibitem[{{Schlegel} {et~al.}(1998){Schlegel}, {Finkbeiner}, \& {Davis}}]{SFD}
{Schlegel}, D.~J., {Finkbeiner}, D.~P., \& {Davis}, M. 1998,
  \href{http://dx.doi.org/10.1086/305772}{\JournalTitle{\apj}, 500, 525}

\bibitem[{{Spiewak} {et~al.}(2022){Spiewak}, {Bailes}, {Miles},
  {Parthasarathy}, {Reardon}, {Shamohammadi}, {Shannon}, {Bhat}, {Buchner},
  {Cameron}, {Camilo}, {Geyer}, {Johnston}, {Karastergiou}, {Keith}, {Kramer},
  {Serylak}, {van Straten}, {Theureau}, \& {Venkatraman
  Krishnan}}]{2022PASA...39...27S}
{Spiewak}, R., {Bailes}, M., {Miles}, M.~T., {et~al.} 2022,
  \href{http://dx.doi.org/10.1017/pasa.2022.19}{\JournalTitle{\pasa}, 39, e027}

\bibitem[{{Stappers} \& {Kramer}(2016)}]{2016mks..confE...9S}
{Stappers}, B., \& {Kramer}, M. 2016,
  \href{http://dx.doi.org/10.22323/1.277.0009}{in MeerKAT Science: On the
  Pathway to the SKA}, 9

\bibitem[{{Stovall} {et~al.}(2015){Stovall}, {Ray}, {Blythe}, {Dowell},
  {Eftekhari}, {Garcia}, {Lazio}, {McCrackan}, {Schinzel}, \&
  {Taylor}}]{2015ApJ...808..156S}
{Stovall}, K., {Ray}, P.~S., {Blythe}, J., {et~al.} 2015,
  \href{http://dx.doi.org/10.1088/0004-637X/808/2/156}{\JournalTitle{\apj},
  808, 156}

\bibitem[{{Verbiest} {et~al.}(2012){Verbiest}, {Weisberg}, {Chael}, {Lee}, \&
  {Lorimer}}]{2012ApJ...755...39V}
{Verbiest}, J.~P.~W., {Weisberg}, J.~M., {Chael}, A.~A., {Lee}, K.~J., \&
  {Lorimer}, D.~R. 2012,
  \href{http://dx.doi.org/10.1088/0004-637X/755/1/39}{\JournalTitle{\apj}, 755,
  39}

\bibitem[{Virtanen {et~al.}(2020)Virtanen, Gommers, Oliphant, Haberland, Reddy,
  Cournapeau, Burovski, Peterson, Weckesser, Bright, {van der Walt}, Brett,
  Wilson, Millman, Mayorov, Nelson, Jones, Kern, Larson, Carey, Polat, Feng,
  Moore, {VanderPlas}, Laxalde, Perktold, Cimrman, Henriksen, Quintero, Harris,
  Archibald, Ribeiro, Pedregosa, {van Mulbregt}, \& {SciPy 1.0
  Contributors}}]{2020SciPy-NMeth}
Virtanen, P., Gommers, R., Oliphant, T.~E., {et~al.} 2020,
  \href{http://dx.doi.org/10.1038/s41592-019-0686-2}{\JournalTitle{Nature
  Methods}, 17, 261}

\bibitem[{{Wang} {et~al.}(2020){Wang}, {Peng}, {Stappers}, {Liu}, {Keith},
  {Lyne}, {Lu}, {Yu}, {Kou}, {Yan}, {Jiang}, {Jin}, {Li}, {Li}, {Qian}, {Wang},
  {Yue}, {Zhang}, {Zhang}, {Zhu}, \& {FAST
  Collaboration}}]{2020ApJ...892...43W}
{Wang}, L., {Peng}, B., {Stappers}, B.~W., {et~al.} 2020,
  \href{http://dx.doi.org/10.3847/1538-4357/ab76cc}{\JournalTitle{\apj}, 892,
  43}

\bibitem[{{Winkel} {et~al.}(2016){Winkel}, {Kerp}, {Fl{\"o}er}, {Kalberla},
  {Ben Bekhti}, {Keller}, \& {Lenz}}]{EBHIS}
{Winkel}, B., {Kerp}, J., {Fl{\"o}er}, L., {et~al.} 2016,
  \href{http://dx.doi.org/10.1051/0004-6361/201527007}{\JournalTitle{\aap},
  585, A41}

\end{thebibliography}

\end{document}